\documentclass[fleqn,usenatbib]{mnras}

\usepackage{newtxtext,newtxmath}
\usepackage{subfigure}  %Subfigures
\usepackage{wasysym} 	%Astronomical symbols
\usepackage{verbatim}	%Block comments

\usepackage[T1]{fontenc}
\usepackage{ae,aecompl}

\usepackage{graphicx}	% Including figure files
\usepackage{gensymb}    % General symbols
\usepackage{amsmath}	% Advanced maths commands
\usepackage{amssymb}	% Extra maths symbols

\title[Moons Around Ejected Giants]{Survivability of Moon Systems Around Ejected Gas Giants}

\author[Rabago and Steffen]{
Ian Rabago$^{1}$ and 
Jason H. Steffen$^{1}$\thanks{E-mail: jason.steffen@unlv.edu}
\\
$^{1}$University of Nevada, Las Vegas\\
}

\date{Accepted XXX. Received YYY; in original form ZZZ}

\pubyear{2018}

\begin{document}
\label{firstpage}
\pagerange{\pageref{firstpage}--\pageref{lastpage}}
\maketitle

\begin{abstract}
%Beginning sentence about planetary formation and escaping planets?
We examine the effects that planetary encounters have on the moon systems of ejected gas giant planets.  We conduct a suite of numerical simulations of planetary systems containing three Jupiter-mass planets (with the innermost planet at 3 AU) up to the point where a planet is ejected from the system.  The ejected planet has an initial system of 100 test-particle moons.  We determine the survival probability of moons at different distances from their host planet, measure the final distribution of orbital elements, examine the stability of resonant configurations, and characterize the properties of moons that are stripped from the planets.  We find that moons are likely to survive in orbits with semi-major axes out beyond 200 planetary radii (0.1 AU in our case).  The orbital inclinations and eccentricities of the surviving moons are broadly distributed and include nearly hyperbolic orbits and retrograde orbits.  We find that a large fraction of moons in two-body and three-body mean-motion resonances also survive planetary ejection with the resonance intact.  The moon-planet interactions, especially in the presence of mean-motion resonance, can keep the interior of the moons molten for billions of years via tidal flexing, as is seen in the moons of the gas giant planets in the solar system.  Given the possibility that life may exist in the subsurface ocean of the Galilean satellite Europa, these results have implications for life on the moons of rogue planets---planets that drift through the galaxy with no host star.

%We examine the survivability of a moon system around a planet that is ejected from its parent star. Planetary systems are simulated using the REBOUND integrator and run forward in time until an escaping planet is generated. A moon system is added before the planet's escape, allowing for the survivability of the planet-moon system to be examined. Preliminary results from the simulations show that moons can survive over a range of orbital distances and eccentricities around the escaping planet.

% DrS --- IS THERE ANYTHING YOU CAN SAY ABOUT THE FUNCTIONAL FORM OF THE SURVIVING MOON DISTRIBUTION, AND/OR THE REASONS THAT THEY CAN SURVIVE?

\end{abstract}


%%%%%%%%%%%%%%%%%%%%%%%%%%%%%%%%%%%%%%%%%%%%%%%%%%

%%%%%%%%%%%%%%%%% BODY OF PAPER %%%%%%%%%%%%%%%%%%

\section{Introduction}
\label{sec:intro}

%Change the citation reference names to make them more recognizable?
Of the locations where we imagine finding evidence for life in the solar system, only one is in the traditional ``habitable zone'' where liquid water exists on the surface due to solar radiation.  Another promising place to imagine finding life is on water-rich moons of the giant planets---with Europa being the most prominent \citep{Squyres:1983,Sparks:2017}.  These moons do not reside (and likely have never resided) within the canonical habitable zone of the Sun.  The energy that keeps their subsurface water in liquid form comes from tidal flexing due to the moons' orbital eccentricity and their relatively close proximity to their host planet \citep{Vance:2018,Reynolds:1987}.  The conditions that provide the energy needed to sustain the liquid water layers have persisted for billions of years in the solar system and are expected to continue for billions more \citep{Yoder:1981}.  Thus, when considering the possibility for life outside the solar system, the moons of gas giant planets are a plausible candidate location \citep{Hinkel:2013,Forgan:2013,Lammer:2014,Zollinger:2017}.

%Lead into this paragraph more
During star formation, systems frequently produce multiple gas giant planets, as seen by Doppler surveys \citep{Knutson:2014,Schlaufman:2016}.  Once the protoplanetary disk dissipates, many of these systems will be unstable.  Dynamical interactions among the planets would lead to the ejection of one or more planets from the system.  Indeed, this scenario is a prominent theory for the formation of hot Jupiter systems \citep{Rasio:1996,Chatterjee:2008} where the encounter that ejects one gas giant simultaneously leaves the remaining planet on a highly eccentric orbit---which then circularizes under the dissipative effects of tidal flexing \citep{Goldreich:1966}.  The final orbit will be at a distance one to two times the original pericenter distance (from conservation of angular momentum while the orbital energy dissipates).

%FIXME: Ian, a lot of your sentences are really, really long with lots of commas.  Those sentences are often hard to parse.  In the future, try to keep your sentences trimmed down (not to the level of "see spot run" but shorter than what you currently have).

If this mechanism does produce the hot Jupiter population, or a sizeable portion of it, it would simultaneously generate a population of ``rogue planets'', ejected from the system to wander the galaxy without a stellar host.  There is some evidence from microlensing searches for the existence of such rogue planets---perhaps numbering in the billions \citep{Sumi:2011,Bennett:2014}.  If these rogue planets can retain their moons throughout the dynamical encounters that lead to their ejection from their home system, then some of those moons may bring with them substantial reservoirs of liquid water.  As is the case in the solar system, the liquid state of this water may persist for billions of years through tidal interactions with the host planet.  Thus, if the moons can survive, it may be possible for life to exist around these planets even in the absence of radiation from a nearby star.

In this work, we use a suite of N-body simulations to estimate the probability of moons surviving in orbit around ejected gas giant planets, and examine some of their anticipated orbital properties.  The paper is organized as follows.  In Section \ref{sec:setup} we detail 77 numerical simulations involving dynamically unstable gas giant systems, and then examine the results of those simulations in Section \ref{sec:results}.  We briefly compare our results with those of \citet{Hong:2018}---a related study that we became aware of as we were preparing this work.  Finally, we discuss some of the broader implications of our results and give our conclusions.

\section{Simulation Setup}
\label{sec:setup}

Our initial systems comprise three Jupiter-mass planets orbiting a solar-mass star.  We assume each planet has a Jupiter radius $R_J$, although this will not play a role in the integration.  The innermost planet is assigned a semi-major axis of 3 AU.  The semi-major axes of the remaining planets are assigned by assigning the orbital period of each planet to be a random ratio with its interior neighbor.  The random ratios are uniformly distributed between 1.2 and 1.4.  The eccentricities $e$ and orbital inclinations $i$ (in radians) of the three planets are drawn from a Rayleigh distribution with a Rayleigh parameter of 0.01.  The longitude of pericenter $\varpi$, longitude of ascending node $\Omega$, and the mean anomaly $M$ are all uniformly distributed between 0 and 2$\pi$.

%Exoplanet solar systems are created and simulated using the REBOUND code (citation). This particular simulation code was chosen for its Simulation Archive feature, which allows for the recording and reinitialization of any particular simulation.
%The simulated systems start as three-planet systems, with a 1 solar mass star in the center surrounded by three Jupiter mass planets. 
%\begin{equation}
%	x=\sigma\sqrt{-2\ln{(1-U)}}
%    \label{eq:rayleigh}
%\end{equation}
%Does not explain why inclination uses Rayleigh instead.
%with the parameter $\sigma$ = 0.01. The other three orbital parameters $\omega$, $\Omega$, and \textit{M} are uniformly distributed.

%Other citation needed for the IAS15 integrator?
%Note that collisions are not implemented.
Once the initial setup is complete, we integrate the systems using the IAS15 adaptive integrator in the REBOUND software package \citep{Rein:2012,Rein:2015}.  The systems evolve until either a planet is ejected or until a maximum time of 10 Myrs is reached.  Our criterion for planet ejection is when the planet reaches an orbital distance greater than 100 AU.  Most systems eject a planet very quickly due to the close proximity of the neighboring planets.

%Moon-planet distance: Use distance or semi-major axis?
% It is not clear why the system is backtracked a certain amount of time. Should the behavior of a vs. t of the planet be mentioned?
Once a planet is ejected from its system, the planet in question is identified and the simulation is restarted at an earlier time using the Simulation Archive feature of REBOUND \citep{Rein:2017}.  The time chosen for the restart is 10 million days prior to the final ejection time (roughly $\frac{1}{300}$ of the maximum allowed simulation time) or from the beginning if the simulation time is shorter.  We add a set of 100 massless moons in circular orbits around the identified planet with the first moon at a distance of $2R_{J}$ from the planet (about one third the orbital distance of Io around Jupiter).  Each subsequent moon is placed an additional $2R_{J}$ outwards.  This creates a disk of moons around the planet reaching out to $\sim 200 R_J$, or roughly 0.1 AU from the planet.  This distance sits well within the planet's initial Hill radius of $\sim 0.7$ AU, which ensures the initial moon orbits are stable.  (In this paper, we will not refer to the moon orbital distances in terms of the Hill radius because the Hill radius will change throughout the simulations as the orbits of the planets evolve.)  This disk of moons is introduced with no inclination relative to the Cartesian coordinate system used by REBOUND---generally placing the disk at a slight angle relative to the planet's orbit.%  No other orbital elements of the moons are modified.
%FIXME: HOW DO YOU CHOOSE THE START TIME FOR THE MOONS IN THE SIMULATION.  IF IT IS TOO CLOSE TO THE FINAL ENCOUNTER, THEN SOME OF THE MOONS THAT WOULDN'T OTHERWISE SURVIVE WOULD ARTIFICIALLY STILL BE IN PLACE.

We note that our initial plan was to use REBOUND's Simulation Archive to put the moons in place assuming that the planetary orbits would remain unchanged.  However, the addition of the moons into the system forced the integrator to adjust its timestep to a smaller value, which caused the orbits of the planets to diverge from their moonless orbits.  This change in timestep caused our escape rate to be less than 100 percent.  Nevertheless, this method still produces a sufficient fraction of escaping systems.  Thus, we continued with this approach and analyzed the planetary escapes that resulted---examining the orbits of the moons that remain around the ejected planet, as well as the fate of moons that were ionized from the planet but that remain in the stellar system.  %Since the host planet is the planet that escaped previously, it is likely that many of these planets were already on a path towards escape before the timestep change was implemented.

%After the moons are added, the simulation is run forward in time again to the point of the planet's ejection from the star system. However, the close distance between the planet and its orbiting moons forces the IAS15 integrator to run at a smaller timestep than usual. This causes the simulation to propagate in a different manner than it did originally, which often leads to the planet no longer escaping. Thus, only a small percent of the original simulations that are run actually create a true escape.

%Once a simulation is completely finished, the last snapshot of the simulation is checked for the number of moons that are still bound to the planet, and various orbital elements of the surviving moons are recorded and analyzed.

\section{Results}
\label{sec:results}
%Percentages will change. Comment on percentage.

%IAN, REMEMBER NOT TO USE PASSIVE VOICE.  "DATA WAS COLLECTED" should be "WE COLLECTED DATA" or something.  DJS

%Figure 1 - Bound Moons
\begin{figure}
\includegraphics[width=\columnwidth]{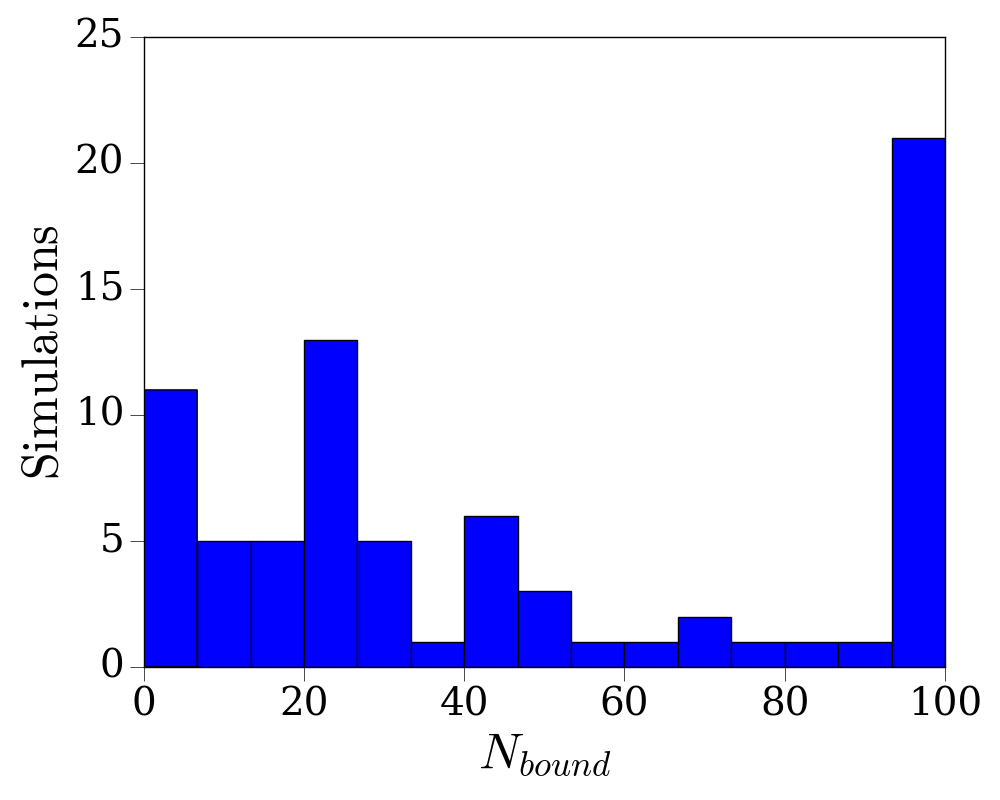}
\caption{Distribution of the number of moons that remain bound to the ejected planet, $N_{bound}$.  Most simulations retain a few to several moons after ejection while just over 25\% of the simulations retain all or most of their moons.  Systems that retain the majority of their moons likely had retrograde encounters with the perturbing planet, thus shortening the duration of the interaction and limiting the change in the moon's velocity.}
\label{fig:moon_bound}
\end{figure}

%FIXME: THE LABELS ON ALL FIGURES ARE WAY TOO SMALL.  THE TICK MARK FONT NEEDS TO BE MUCH, MUCH LARGER.

%Figure 2 - Surviving moon distribution.
\begin{figure}
\includegraphics[width=\columnwidth]{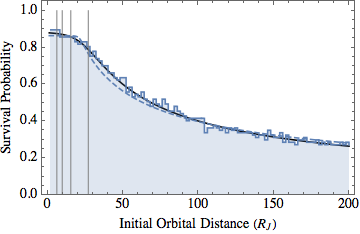}
\caption{Fraction of surviving moons as a function of the moon's initial distance from the parent planet.  The vertical lines indicate the orbital distances of the Galilean moons, where a large majority of moons survive.  The results of two, broken power-law fits to the data are shown.  The blue dashed line had a fixed power-law index of $-1/2$.  The solid black line is an improved fit with an index of roughly $-0.6$.}
\label{fig:moon_dist}
\end{figure}

%Figure 3 - Histograms of a, e, and i.
\begin{figure*}
	\centering
	\subfigure[]{\includegraphics[width=0.65\columnwidth]{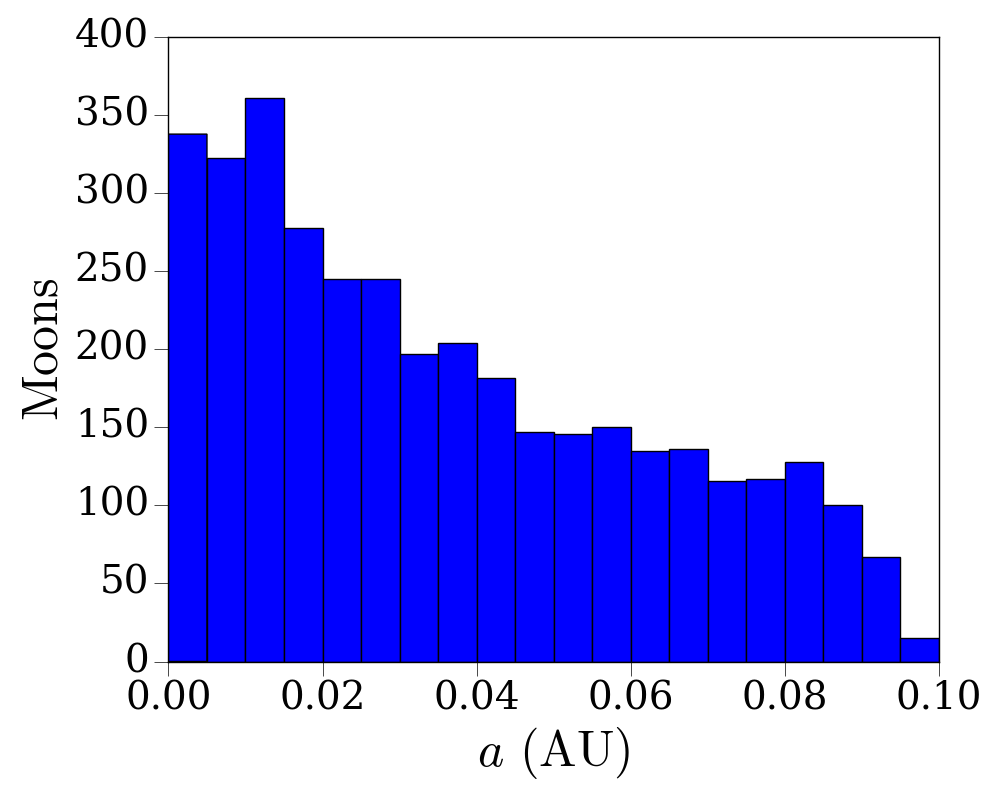}\label{fig:moon_a}}
    \subfigure[]{\includegraphics[width=0.65\columnwidth]{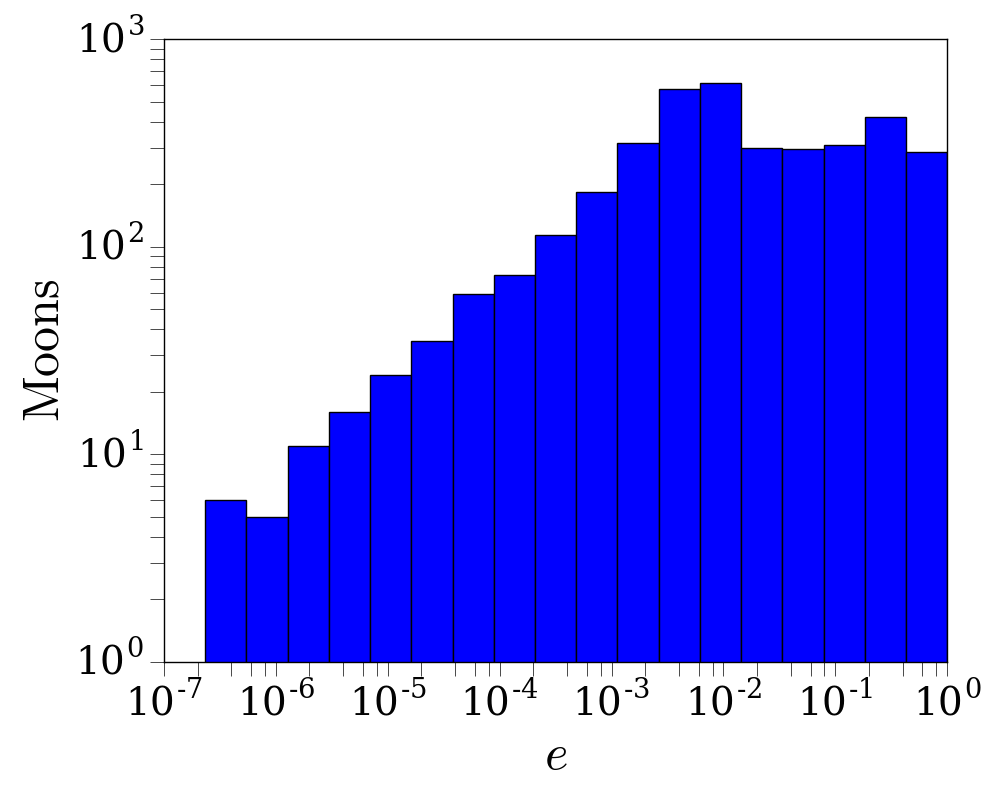}\label{fig:moon_e}}
    \subfigure[]{\includegraphics[width=0.65\columnwidth]{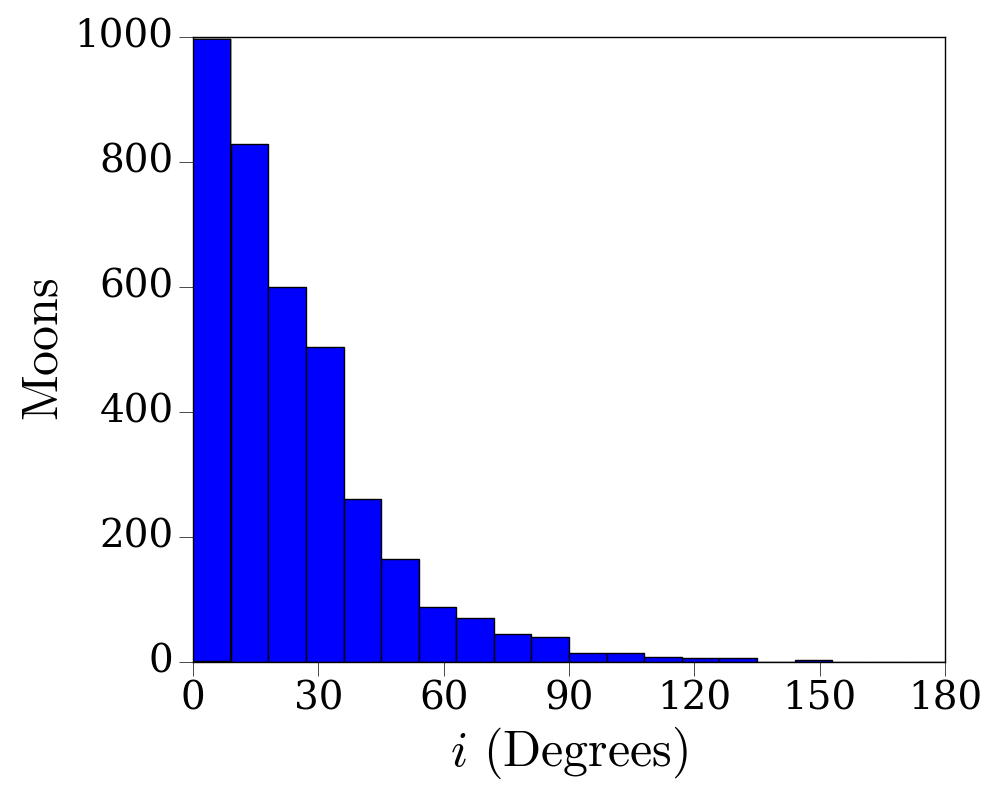}\label{fig:moon_inc}}\\
    
	\subfigure[]{\includegraphics[width=0.65\columnwidth]{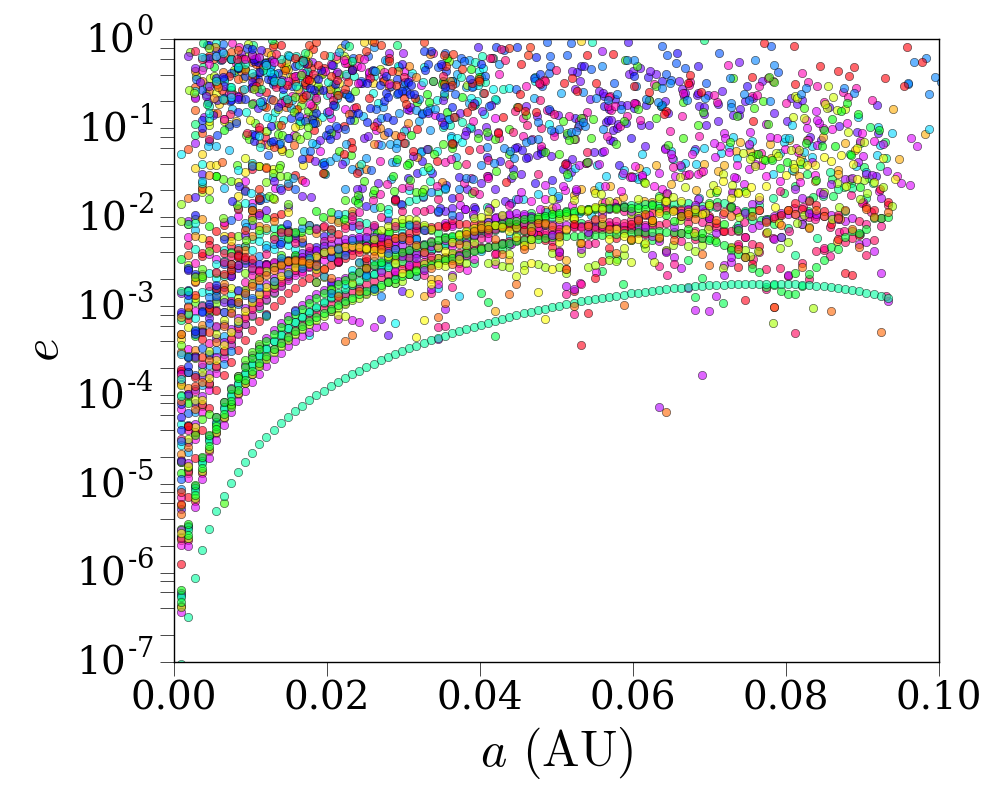}\label{fig:moon_ae}}
    \subfigure[]{\includegraphics[width=0.65\columnwidth]{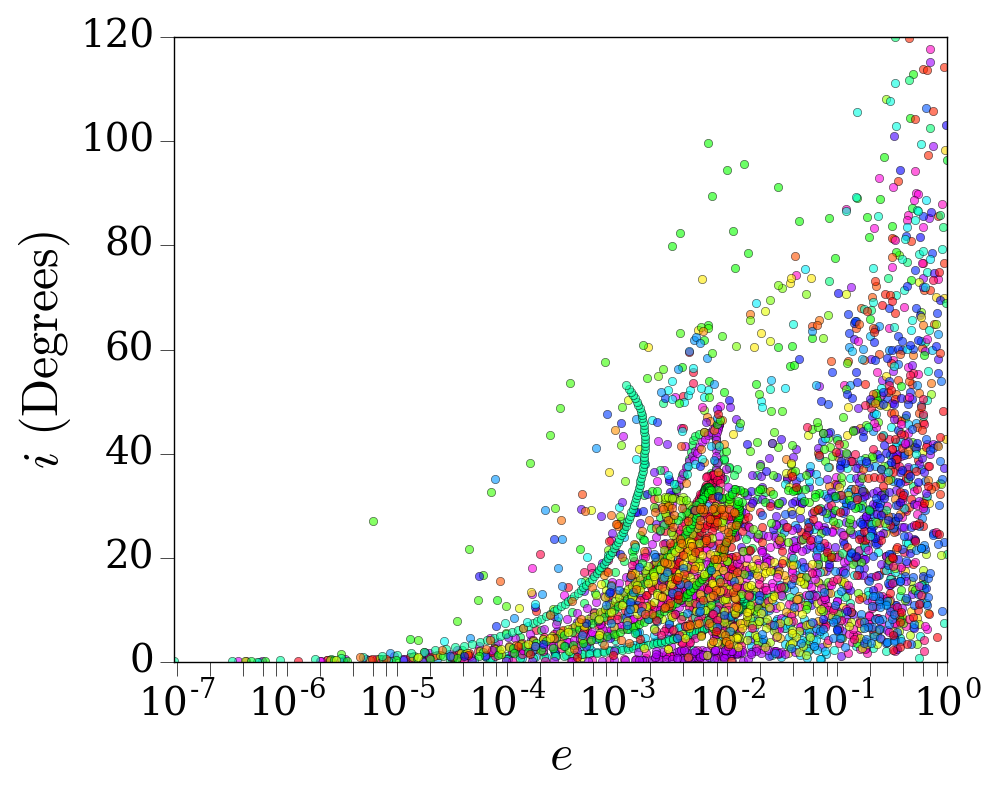}\label{fig:moon_ei}}
    \subfigure[]{\includegraphics[width=0.65\columnwidth]{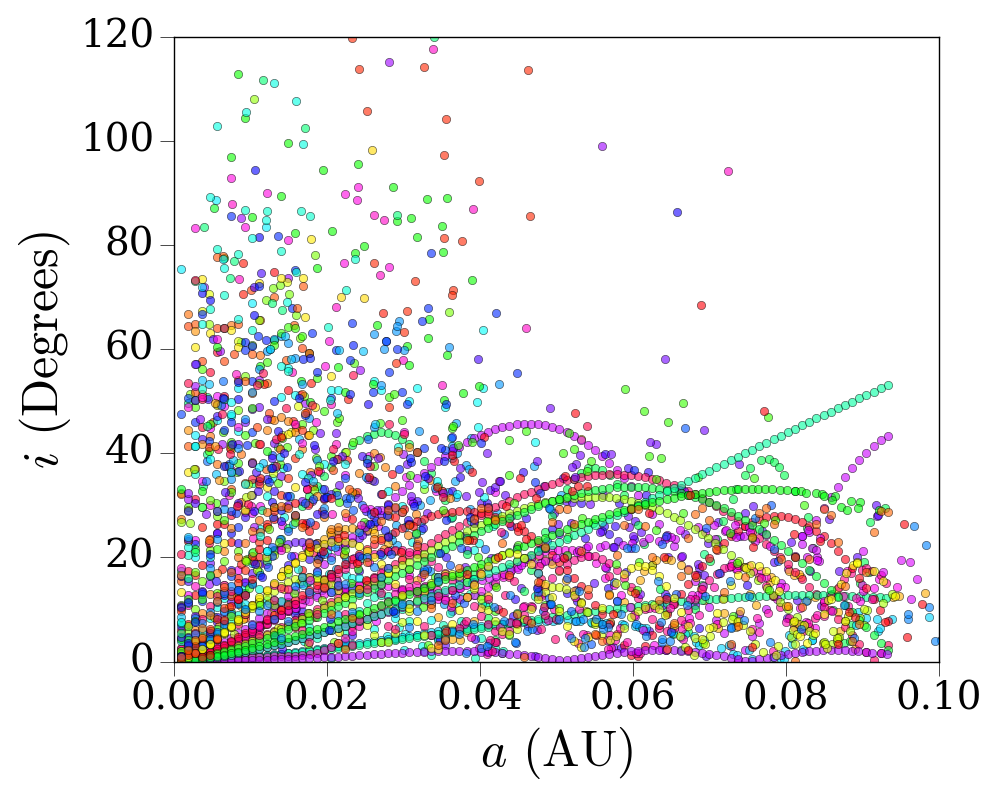}\label{fig:moon_ai}}
    
    \caption{Top Row: The distributions of the final orbital elements (\textit{a}, \textit{e}, and \textit{i}) of the moons that survive the planetary ejection.   Eccentricity is plotted on a log-log scale.  Bottom Row: Scatter plots of pairs of final orbital elements of the moons that survive the planetary ejections.  In both sets of figures we see that the eccentricity distribution of the final moons has two clusters of points, one near $10^{-2}$ composed largely of the systems that retain the majority of the moons and the second with eccentricities of a few tenths.  The coherent structures in the bottom panels are the result of precession in the moon orbits as the planet is perturbed onto an inclined orbit prior to being ejected from the system.}
\label{fig:moon_1d}
\end{figure*}

%Figure 4 - 2D plots of a, e, and i
%\begin{figure*}
%	\centering
%	\subfigure[] {\includegraphics[width=0.6\columnwidth]{Moon_ae_log}\label{fig:moon_ae}}
%    \subfigure[] {\includegraphics[width=0.6\columnwidth]{Moon_ei_log}\label{fig:moon_ei}}
%    \subfigure[] {\includegraphics[width=0.6\columnwidth]{Moon_ai}\label{fig:moon_ai}}
    
%    \caption{2D plots of \textit{a}, \textit{e}, and \textit{i} of surviving moons, collected over all simulations. Eccentricity is plotted on a log scale.  Of note are the patterns present in each graph---these are discussed in detail in Figure~\ref{fig:inc_Omega}.}
%    \label{fig:moon_2d}
%\end{figure*}

%FIXME: what are the chances of getting up to 100 simulations that produced the ejected planets with moons---that would be best.

We ran a total of 77 simulations with escaping planets over the course of this experiment and found that 47\% of the moons remain bound to the escaping planets at the end of the simulation.  An additional 22\% of the moons were stripped from the planet, but remained bound to the star.  The remaining 31\% were stripped from both the planet and the star.

Figure~\ref{fig:moon_bound} shows the distribution of the number of moons that were retained by the escaping planet.  We see from this figure that the distribution of orbital distances of surviving moons is bimodal.  Figure~\ref{fig:moon_dist} shows the survival rate for the moons as a function of the moon's initial distance from the planet.  In general, the scattering events strip the outermost moons from the system and most systems lose some moons.  However, a substantial fraction of the systems retain all of their moons.  (Recall that these are non-interacting, test particle moons.  A physical system of 100 interacting moons is unlikely to survive unscathed).  Also shown in Figure \ref{fig:moon_dist} are the locations of the Galilean satellites of Jupiter.  We see from this figure that a large fraction of the moons ($\sim 85$\%) near the orbits of the Galilean satellites will survive the ejection of the planet from the system.

%FIXME: INSERT PARAGRAPH HERE ABOUT THE SHAPE OF THE DISTRIBUTION, ALSO NOTE THAT THE SHAPE LIKELY DEPENDS UPON THE MASS OF THE SCATTERING PLANETS.

The overall shape of the distribution in Figure \ref{fig:moon_dist} can be modeled as a broken power law with a constant value for small semi-major axes and a power-law tail for larger values.  The tail scales approximately like $a^{-1/2}$, however, a far superior fit to the data occurs when the power-law index is left as a free parameter.  A least-squares fit to the logarithm of the survival probability as a function of the logarithm of the semi-major axis yields an index of $-0.607 \pm 0.017$.

While many of the moons survive after the planet ejection, their orbits are often significantly disrupted.  Figures ~\ref{fig:moon_a}, ~\ref{fig:moon_e}, and ~\ref{fig:moon_inc} show the distributions of the semi-major axes, eccentricity, and inclination for the surviving moons, while Figures ~\ref{fig:moon_ae}, ~\ref{fig:moon_ei}, and ~\ref{fig:moon_ai} show 2-dimensional plots of these elements.  In these figures, moons are shown at their final orbital configuration with orbital elements calculated in reference to the host planet.  We show only moons that have remained bound to the planet.

A few features stand out in these figures.  A large fraction of the surviving moons have nearly circular orbits with the remainder of the eccentricities spread throughout the allowed range.  The distribution of semi-major axis of the remaining moons also spans the entire range of the initial conditions.  The orbits of the moons are reordered somewhat (as can be seen by comparing the final distribution in Figure \ref{fig:moon_a} to the initial distribution in Figure \ref{fig:moon_dist}) but most moons stay relatively close to their initial orbits.  The final orbital inclinations are generally modest but the distribution is quite wide and extends to both polar and retrograde orbits in the most extreme cases.  We assume the inclinations are often excited because the planets are first scattered into inclined orbits before being ejected from the system---which initiates stellar-induced changes to the inclination of the moon systems.

Another striking feature is the occasional, coherent pattern in the inclination of the surviving moons from several of the ejected planets.  Systems that show these patterns generally retain most or all of their moons and can be identified in Figure~\ref{fig:moon_1d} by observing distinct patterns to their final moons.  These structures are caused by the precession of their orbits induced by the central star.  Once a a planet is scattered out of the plane of its initial orbit, the orbits of all of the moons become misaligned from the orbit of the planet around the star.  The gravitational interaction between the star and the moons causes them to precess at different rates---producing those patterns.  In some cases, the patterns develop in initial scattering events, but the subsequent planet-planet encounters eventually cause the planetary ejection will remove multiple moons---leaving partial precession patterns or disrupting them completely.

\begin{figure}
\includegraphics[width=\columnwidth]{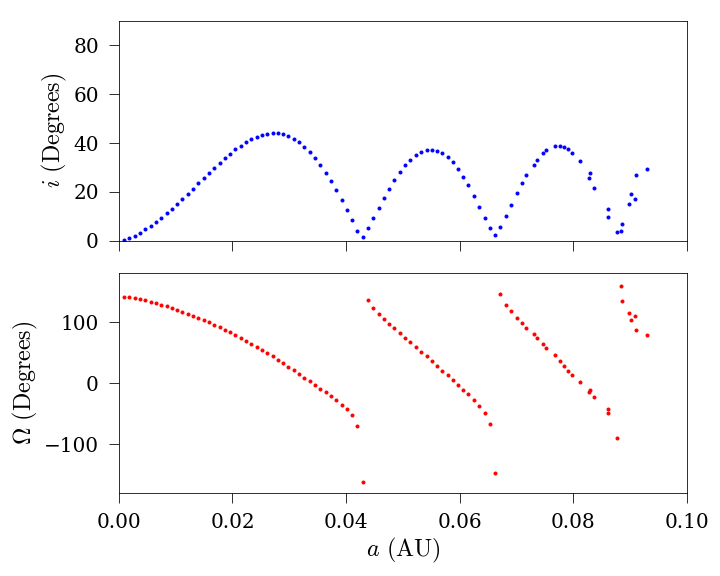}
\caption{Graph of inclination and longitude of ascending node for moons in a simulation that retained all of its moons.  Orbital elements are taken $10^6$ days after the moons are first added. Points of $0\degree$ inclination (blue) coincide with large changes in $\Omega$ (red) where the orbital elements cycle from $-180^\circ$ to $+180^\circ$.}
\label{fig:inc_Omega}
\end{figure}

This pattern in inclination for an example system is shown in Figure~\ref{fig:inc_Omega}, along with the longitude of ascending node $\Omega$ for the moon system.  In this figure, the full precession structure can be seen.  Large changes in the value of $\Omega$ indicate where its value completes a circuit of $2\pi$ and where its inclination is $0\degree$.%  This observation suggests the orbits precess back and forth around $0\degree$ inclination.  As the inclination of the orbit crosses $0\degree$, the ascending part of the orbit moves from one side to the other, causing a large change in the value of $\Omega$.

We do not explore the cause of the high retention rate in these systems in this work.  However, we suspect that the systems that retain the majority of their moons experienced scattering events that were retrograde with respect to the moon orbits.  Since the velocity imparted to the moons is proportional to the crossing time of the encounter, it will be inversely proportional to the relative velocity of the moon and the perturber:
\begin{equation}
\Delta v_{\text{moon}} = \frac{F_{\text{G}}}{m_{\text{moon}}} \Delta t \simeq \frac{G m_{\text{p}}}{b \Delta v}
\end{equation}
where $\Delta v_{\text{moon}}$ is the velocity imparted to the moon, $F_{\text{G}}$ is the force of gravity (with Newton's constant $G$), $\Delta t$ is the passage time, $m_{\text{p}}$ is the mass of the perturbing planet, $b$ is the distance of closest approach (also assumed to be the crossing distance), and $\Delta v$ is the difference between the velocities of the moon and the perturbing planet.  In a prograde encounter, the moon and perturber are traveling in the same direction, yielding a small relative velocity, a long crossing time, and a large change to the velocity of the moon.  In a retrograde encounter, the moon is traveling in the opposite direction, producing a large $\Delta v$, a small crossing time, and a small perturbation to the moon's velocity.  We inspected several systems where all of the moons were retained and each had primarily retrograde encounters---supporting our hypothesis.  Nevertheless, a comprehensive exploration of this issue may prove interesting.%  An alternative explanation is that the group of systems with 100\% retention simply reflects all of the systems that would lose moons if their orbits extended beyond the outer limit of 200 Jupiter radii that we impose from our initial conditions (though this explanation seems unlikely to account ).

\begin{figure}
	\centering
	\subfigure{\includegraphics[width=0.95\columnwidth]{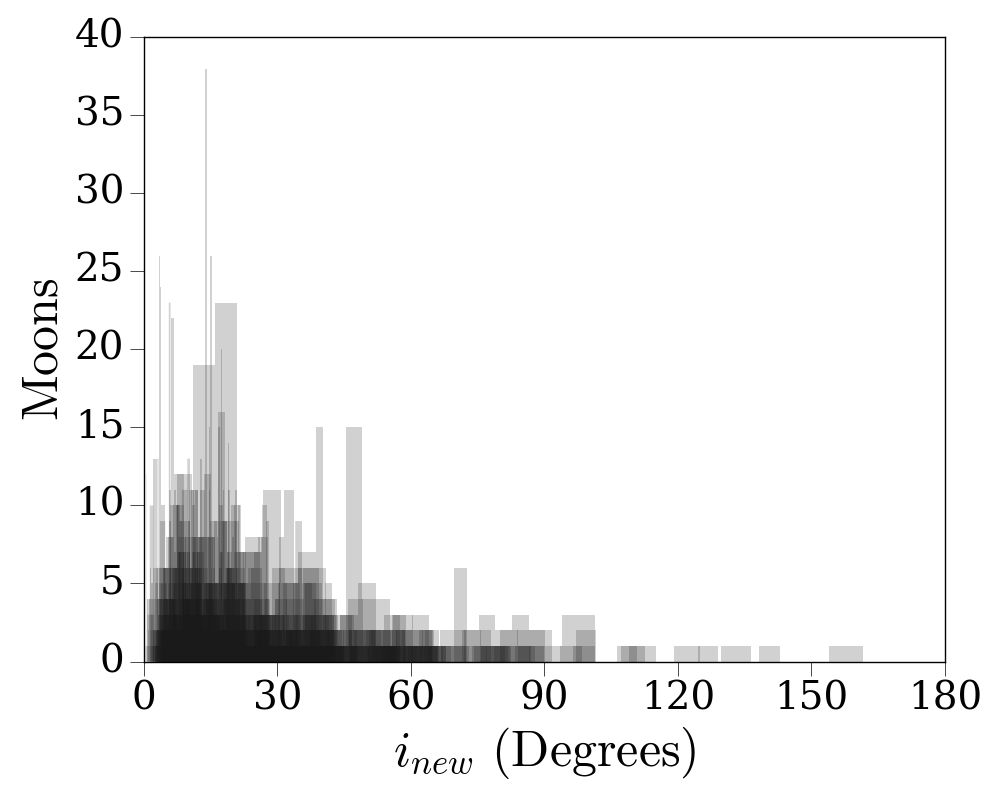}\label{fig:newinc_ind}}
    \subfigure{\includegraphics[width=0.95\columnwidth]{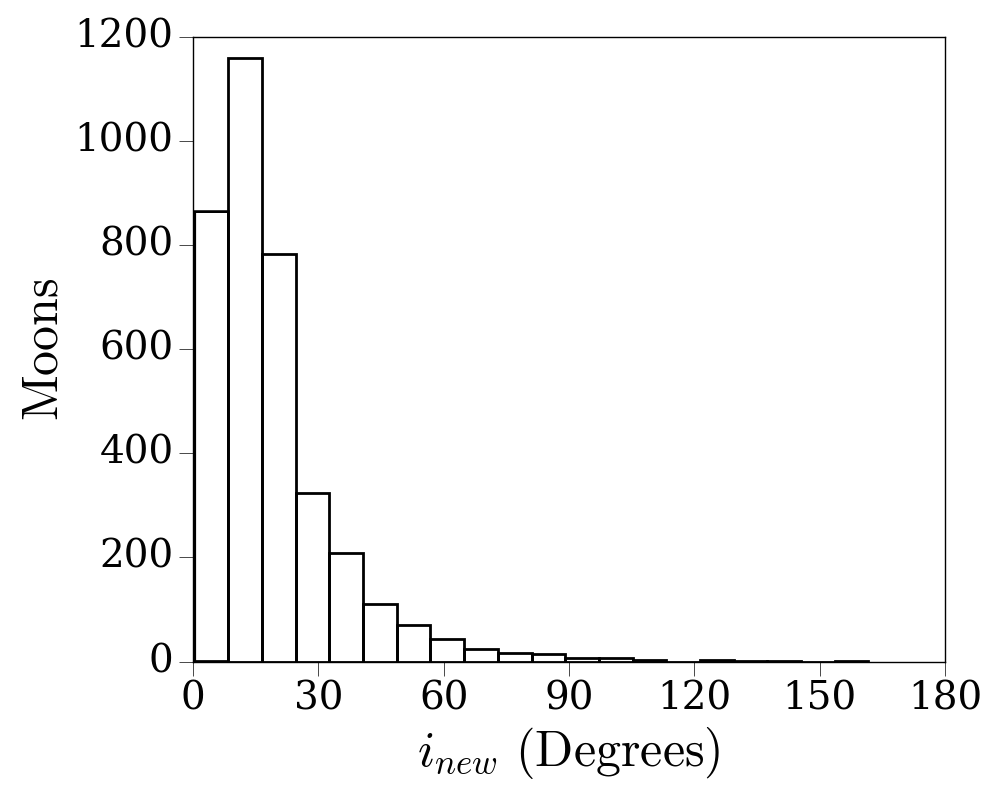}\label{fig:newinc_total}}
    
    \caption{Inclination distributions of surviving moons with respect to their common orbital plane (defined by the direction of the total angular momentum of the moons).  Top: Distributions of individual simulations.  Darker regions correspond to values that are densely occupied by surviving moons.  Bottom: The combined distribution from data collected over all simulations.  The peak inclination is distinctly nonzero in these distributions.}
    \label{fig:newinc}
\end{figure}

%Check actual inclination distribution - is it enough to prevent resonances?
%For citation, see Gallardo et. al. (2016)
%ArXiV 1603.06911
We note that the distribution of \textit{i} measured in Figure~\ref{fig:moon_inc}---as well as in the preceding paragraphs---is created in reference to the initial coordinate system of REBOUND.  However, as the inclinations of the moons in the moon system vary over time we can consider how they are distributed about a mutual orbital plane.  We define the mutual orbital plane to be the plane perpendicular to the total angular momentum vector of all of the moons that remain bound to the planet.  We recalculate the inclination distribution with respect to this new orbital plane and show it in Figure~\ref{fig:newinc}.  Distributions of individual systems are also shown in that figure, where a variety of spreads are seen from system to system depending upon the angle between the initial moon disk and the orbit of the host planet prior to its ejection.  Figure~\ref{fig:newinc} shows the inclination distribution when combined over all simulations and suggests that, when measured against their common orbital plane, moons are likely to have a nonzero value for $i_{new}$.  Examining individual simulations shows that $i_{new}$ is nonzero even at small orbital distances from the planet.%This angle is the same as the planet's inclination as measured to the Cartesian system. The peak of the histogram in Figure~\ref{fig:inc_new} is shifted rightwards compared to the peak in Figure~\ref{fig:moon_inc}. 

Ultimately, however, the visual aspect of these patterns are an artifact of our simulation setup since real moon systems are not this regular in their orbital properties---and the moons are not massless.  One could imagine seeing the effects of such precession in a hypothetical, if unlikely, statistical study (of moons orbiting rogue planets or planets whose orbit are highly inclined with respect to other planets in the same system).  Other effects that we do not consider here, such as planet oblateness, moon mass, and formation processes, will also contribute to the architecture of the surviving moon systems.  Thus, our distributions of inclination, eccentricity, and semi-major axis are best viewed as the parameter space where the orbits of individual moons would be found and not necessarily how moons with in a particular system would relate to each other.  Nevertheless, we expect the bulk properties of the surviving moon distribution to roughly match these results.

An interesting question to consider on similar statistical grounds is whether or not resonant orbits can survive the ejection process.  Mean-motion resonances are common among moon systems in the solar system, and would presumably be common in moon systems of extrasolar planets \citep[e.g.,][]{Murray:1999}.  Thus, we examine the effects that planet-planet scattering has on such resonant dynamics (now using massive moons) in section \ref{sec:resonance}.

%The distribution shown in Figure~\ref{fig:moon_bound} is somewhat bimodal, with a large peak sitting at $N_{bound} = 100$.  This peak represents a population of systems that keep all of their moons during the process of planetary ejection. In these simulations, the eccentricity of these moons is very near 0, but the inclinations of these moons vary widely with distance, increasing to almost $45 \degree$ for some simulations.  Examining these systems individually over time shows these configurations arise as a result of repeated interactions between the moons and the central star.  The gravitational force from the star generates a torque on the disk, which slightly increases the inclination of each moon in the disk. This effect is stronger for moons which orbit at farther distances, and is at its strongest as the host planet passes through perihelion, when the gravitational force from the star is the strongest. Over several passes, the inclination of the orbits oscillates between $0 \degree$ and twice the inclination of the planet's orbit relative to the star. In addition, the orbits change orientation in space, precessing in the variable $\Omega$ over time. 

%Simulations with a partial number of remaining moons often exhibit this effect, but the stronger short-term effects from planet-planet interactions produce large scattering in the disk, which makes observation of this effect difficult in partially surviving systems.

%Figure 5 - Graph of i and Omega for one system.

%Figure 6 - Recalculated Inclination Distribution

\section{Moons that are left behind}

\begin{figure*}
	\centering
	\subfigure[]{\includegraphics[width=0.65\columnwidth]{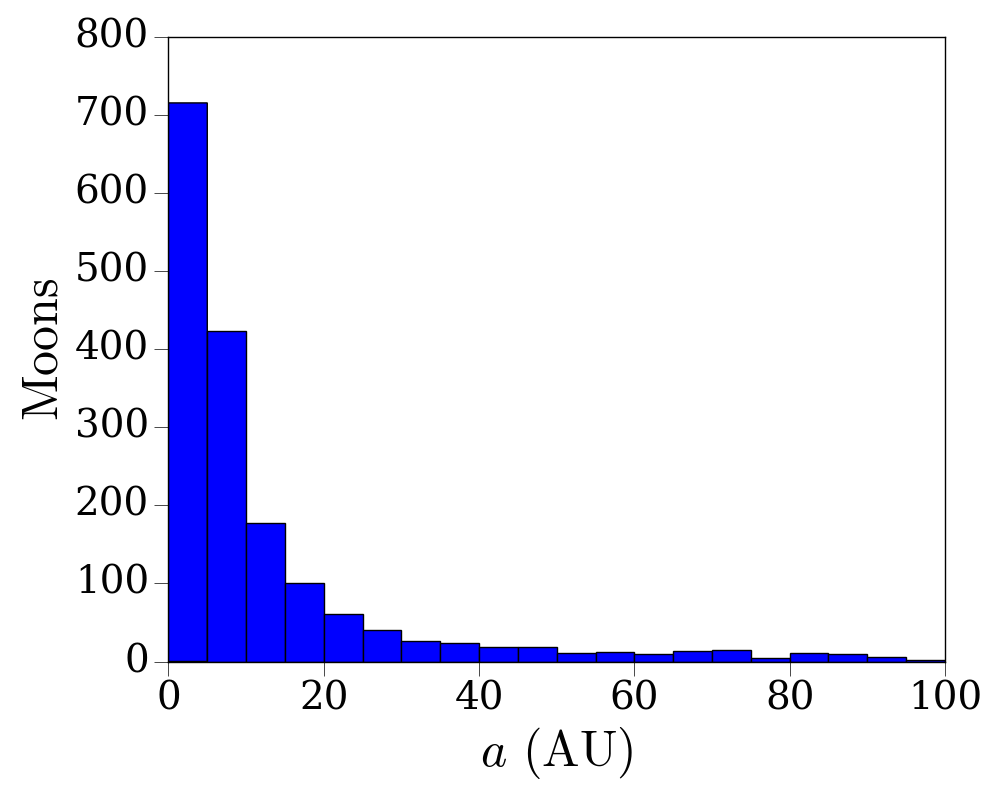}\label{fig:free_a}}
    \subfigure[]{\includegraphics[width=0.65\columnwidth]{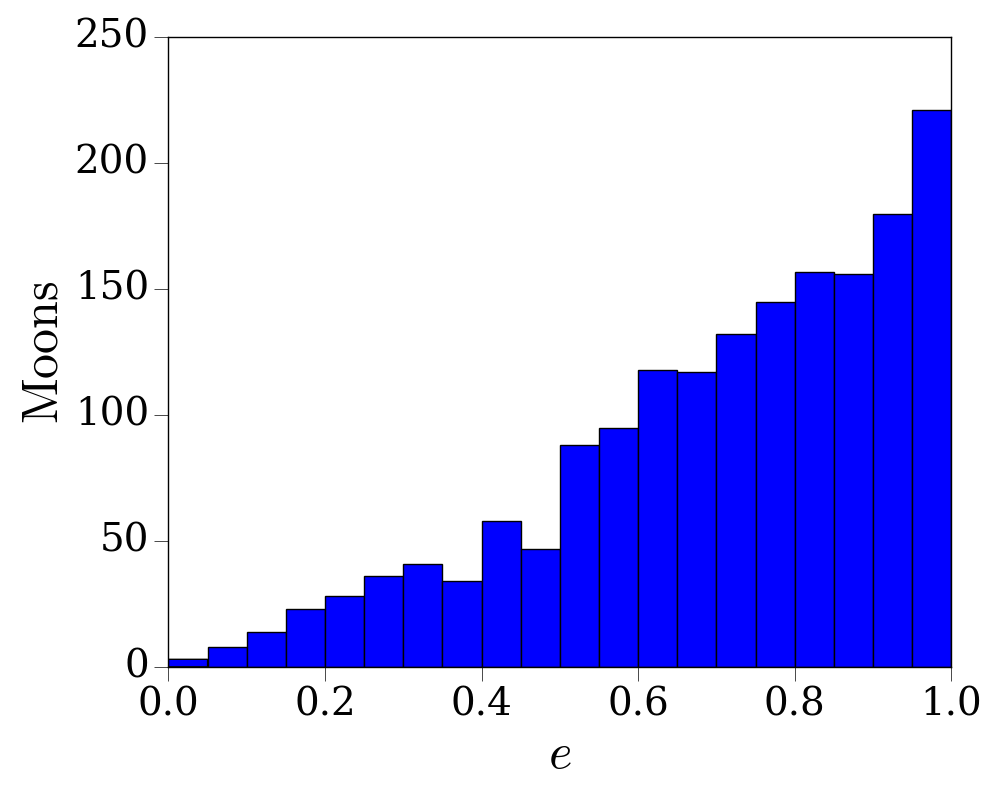}\label{fig:free_e}}
    \subfigure[]{\includegraphics[width=0.65\columnwidth]{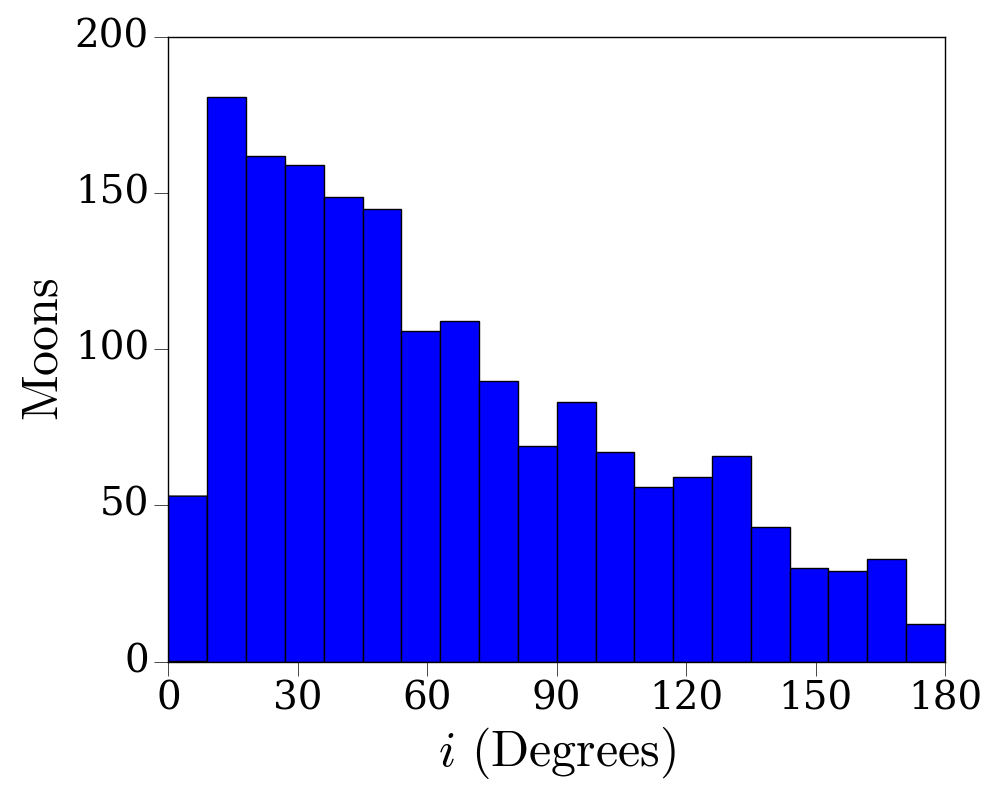}\label{fig:free_inc}}
    
    \caption{Distribution of \textit{a}, \textit{e}, and \textit{i} for moons that are unbound from their host planet, but remain bound to the central star.  Elements are calculated with respect to the system's center of mass, and with respect to the simulation's initial coordinate system.  The majority of these remnant moons have small orbital distances (compared with our ejection criterion of 100AU).  These moons favor large values of eccentricity and inclination---including a large fraction of polar and retrograde orbits.}
    \label{fig:free}
\end{figure*}

%Possibly analyze a graph of e vs. i to see if a relation exists between the two.
Our systems produced a large population of moons that are stripped from their host planet, but that remain bound to the star.  Figure~\ref{fig:free} shows the orbital elements of these moons (calculated with respect to the initial coordinate system of the simulation).  Figure~\ref{fig:free_e} shows a correlation in the eccentricity distribution---a larger number of moons have high eccentricity.  This feature is, in part, a consequence of the fact that our planetary orbits are at a few AU, while the escape distance is 100 AU.  Thus, the scattering happens close to the Sun (where the future pericenter of the final orbit will be), but the moons can have semi-major axes that are much larger---producing a large apocenter.
%Ambiguous wording

The inclination distribution in Figure~\ref{fig:free_inc} shows that moons span the full range of values from 0 to 180 degrees.  Most moons have only modest inclinations of a few tens of degrees.  Nevertheless, roughly 30\% of the simulated moons have an inclination at or above 90 degrees---occupying polar or retrograde orbits around the star.%We suspect this distribution reflects the angle at which the moon is scattered during a planet-planet encounter. Figure~\ref{fig:free_inc} suggests there is only a slight preference to the angle \textit{i} in the prograde direction $(0 < i < 90)$. Thus, unbound moons likely exist in many different orbital configurations around the star.

%Figure 6 - Unbound Moons

\section{Mean-Motion Resonances}
\label{sec:resonance}

%FIXME: THIS ARTICLE HAS SOME USEFUL REFERENCES
%https://www.sciencedirect.com/science/article/pii/S0019103504001952

%Resonance paragraph. Change depending on current outcomes of resonance.
Another important effect we consider, especially when imagining the potential to harbor life, is whether or not the details of the dynamical state of a moon system remains---specifically the resonance behavior.  Without a perturbing body, the orbit of the moon would circularize, removing an essential ingredient for tidal heating \citep{Yoder:1981}---though a spin-orbit coupling other than 1:1 would allow eccentricity to persist (as is the case with Mercury's eccentricity of 0.2).  And, while nonresonant configurations can induce a forced eccentricity in the orbits of the moons, they would generally be quite small (of order the moon/planet mass ratio) since the conjunctions that produce the eccentricities occur randomly about the orbit.  Either way, the resonant configuration of Jupiter's Galilean moons plays an important role in keeping the eccentricities of Io and Europa large enough to allow tidal heating to warm their interiors \citep{Yoder:1979}.

Capturing two moons into mean-motion resonance beginning from a non-resonant configuration usually requires that their orbital period ratio converges from a larger value to the resonance value \citep{Peale:1976,Lee:2002}.  This situation can arise in the presence of the disk from which the moons form---where the disk dissipates the orbital energy of the outer moon faster than it does the inner moon.  Without a disk it may still be possible to produce a resonant configuration through convergent migration (e.g., by the tidal decay of an exterior moon on an eccentric orbit).  For lunar orbits around a rogue planet, the most straightforward way to generate a resonant configuration is by simply to preserve that configuration throughout the ejection process.

To see if resonances can survive ejection, we replace the disk of moons described in Section \ref{sec:setup} with two different sets of resonant moon systems.  The first set uses two-moon systems in a 2:1 orbital resonance.  The second set uses three moon systems in the three-body, Laplace resonance.  These moons are one lunar mass each, and are positioned at the orbital distances of Io, Europa, and Ganymede from the planet.  Once positioned, the simulation is integrated to the planetary escape as normal.  For all systems, we record the 2:1 resonant argument, $\phi = \lambda_1 - 2\lambda_2 + \varpi_1$, of the inner planet pair as it evolves over the simulation time.  In a sustained 2:1 resonance, this value oscillates around $0\degree$.  For the Laplace resonance systems, we record the 2:1 resonances between the outer pair of moons and the Laplace resonant argument $\Phi_L = \lambda_1 - 3\lambda_2 + 2\lambda_3$, which oscillates about $180\degree$.
%Nevertheless, WE DID SOMETHING?  OR, DID WE JUST KEEP GOING AND ANALYZE WHAT WE COULD?

%FIXME: UPDATE THIS PARAGRAPH WITH NEW RESULTS

We run 10 simulations from each set (the 2:1, two-moon and the Laplace, three-moon configurations), with the moons starting in their respective resonances.  Following the ejection of the planet from the system, the resonant arguments continue to oscillate in seven of the 10, 2:1 resonance simulations.  Eight of the 10 Laplace resonance simulations retain some resonance behavior---six keep the Laplace resonance, one system keeps the individual 2:1 resonances, but not the Laplace resonance, and one keeps the inner 2:1 resonance but not the outer resonance.  These results show that resonances are often retained throughout the ejection process.  These numbers are also consistent with the overall moon survival rate near the Galilean moon orbital distance seen in Figure \ref{fig:moon_dist}.

\begin{figure}
	\includegraphics[width=\columnwidth]{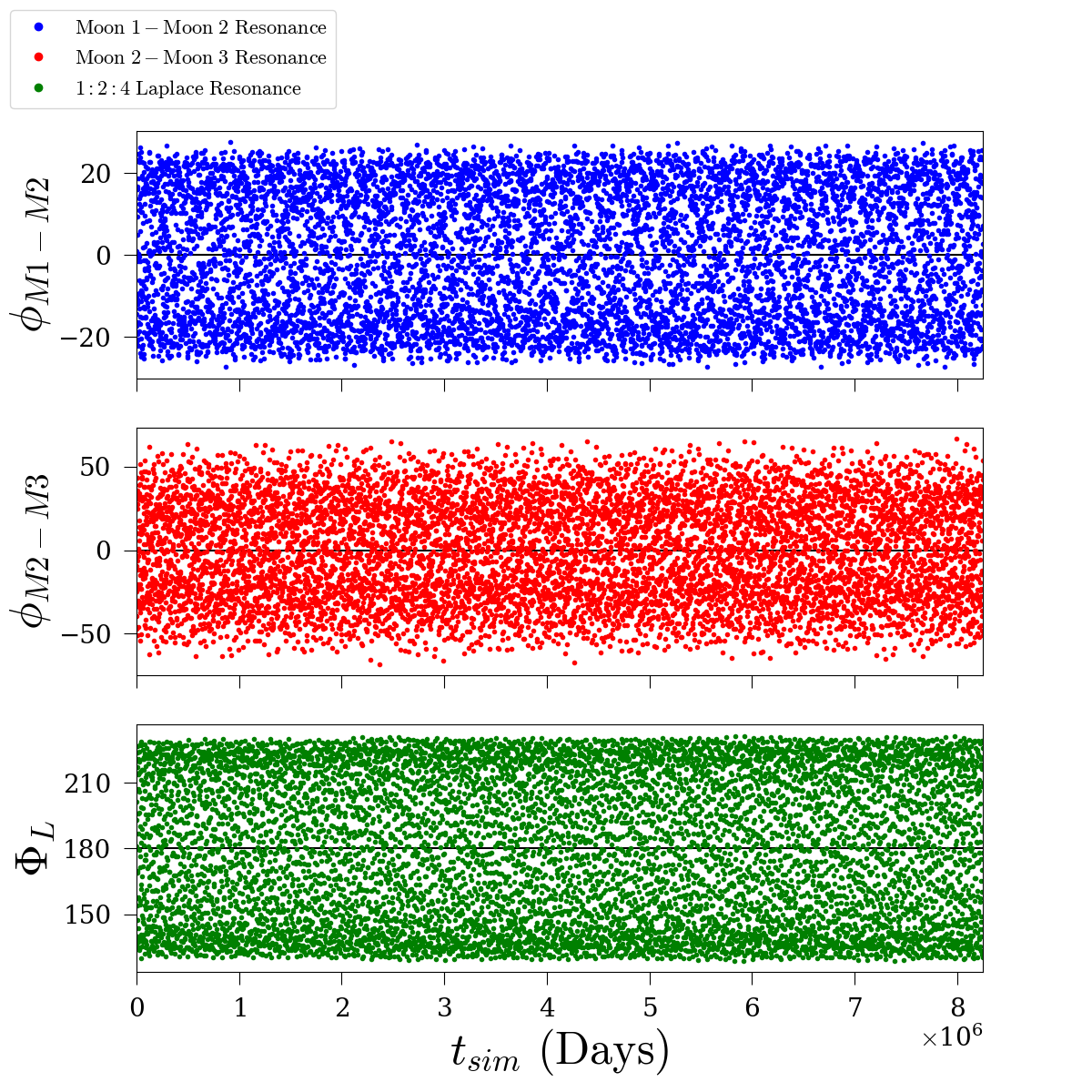}
    \caption{Resonance angles as a function of time for the two 2:1 (top and middle) and the Laplace (bottom) resonances for one of our Laplace resonance, three-moon systems.  The time series spans the entirety of the simulation, including the ejection of the host planet from the system.  This result shows that the resonance behavior of the system can be preserved throughout the evolution of the system and will persist after the host planet leaves the system.  Thus, the conditions on the Galilean satellites (such as volcanoes on Io and liquid water on Europa) could persist on rogue planets for billions of years.}
    \label{fig:resonance}
\end{figure}

As an example, we show the evolution of a Laplace resonance system in Figure~\ref{fig:resonance}, where each of the three resonant arguments is plotted over the entire simulation time ($t_{sim}$).  We see in this figure that the ejection process can have very little effect on the dynamical state of the moon system---even though several encounter events occur before the planet finally leaves the system near the end of the time series.  These simulations demonstrate the potential survival of a resonant moon configuration through the ejection of the host planet---and therefore the potential that such resonant configurations may exist about rogue planets.  Subsequent evolution of the moons' orbits through tidal interactions may restore the Laplace resonance at a future time \citep{Deienno:2014}.

%Is multi-body resonance required for liquid water?

%In these simulations, we find the amplitude of the resonance argument does not change significantly unless the resonance is broken.

%Since the Laplace resonance is created from an interlocked series of 2:1 resonances, there are many ways for it to be disturbed by a planetary encounter.  Of the four simulations created with a Laplace resonance, but where the Laplace resonance is broken, close examination of the remaining moon system reveals one simulation where the individual 2:1 resonances persist without forming a Laplace resonance.  Another simulation is found where the 2:1 resonance between the first two moons survived, but the resonance between the next two did not.  

We also note that moons that stay in resonance also maintain a common orbital plane throughout the simulation---though the orientation of this plane can shift over time.  Some simulations in resonance produce coplanar moon systems with large inclinations by the time the host planet is ejected---up to $60\degree$ from their initial configuration in the largest cases.  For real systems, if the orbits become too inclined, the resonances may be affected by the oblateness of the host planet.

%  FIXME: ALMOST ALL OF THESE NEW SENTENCES ARE IN PASSIVE VOICE.  I'VE CHANGED PROBABLY 6-8 SENTENCES IN THE LAST FEW PARAGRAPHS ALONE.  STOP USING PASSIVE VOICE IN YOUR WRITING.
%Resonances may be slightly out of alignment - can they fall back into alignment?

Our initial set of resonance systems had moons near the orbits of the Galilean satellites---quite close to the host planet where a large fraction of the orbits survive (Figure \ref{fig:moon_dist}).  We also tested resonance configurations in larger initial orbits where lunar ionization is more likely to occur.  For these simulations, we move the inner moon out to either 40$R_J$ or 75$R_J$---giving them a typical survival rate of $\sim 0.7$ and $\sim 0.5$, respectively (as estimated from Figure \ref{fig:moon_dist}).  We then place a second moon outside the first to create a 2:1 resonance.  Ten simulations are run for each orbital distance.  For the 40$R_J$ suite, four of the 10 simulations maintain the 2:1 resonance, while another two simulations retain both moons but lose the resonant interaction.  This results in a total of ten surviving moons out of the initial 20---slightly less than what we would expect from the test particle simulations.  For the 75$R_J$ suite, a total of five moons are retained across all 10 simulations (25\%), but no resonant configurations survive.  Thus, while resonant orbits can survive the ejection process---even on orbits larger than the Galilean satellites---a larger fraction of the moons will be lost.  Moreover, the resonances are (not surprisingly) more fragile than the individual orbits of the moons.

%FIXME: YOU NEED A PARAGRAPH DESCRIBING ALL OF THE RESONANT SIMULATIONS YOU DID---THOSE AT THE HALF-SURVIVAL PROBABILITY DISTANCE AND AT THE 1/4 (OR WHATEVER) DISTANCE.  THAT IS AN IMPORTANT RESULT---SHOWING THAT THE RESONANCE WILL BREAK BEFORE THE MOONS ARE IONIZED, BUT THAT SOME STILL SURVIVE.

% Graphs currently use subfigures to group similar figures. The figures can be made bigger or smaller by changing the width parameter.

%Figure 7 - Resonance Argument

\section{Comparison with other work}
\label{sec:compare}
Our results share similarities with \citet{Hong:2018}, a recent work which also examined the dynamics of moons during close planetary encounters\footnote{We became aware of the independent work of \citet{Hong:2018} as we prepared to present our initial results at the January 2018 meeting of the American Astronomical Society, \citep{Rabago:2018}.}.  We focus primarily on the effects that planet ejection has on the survivability of moons and the evolution of their orbital configuration.  Thus, we only consider the moons that orbit the planet that is eventually ejected from the system and our initial conditions concentrate our moon sample closer to our planet than what was done in \citet{Hong:2018}.  The planetary systems in both works are similar---a star orbited by three planets.  They used a variety of planet masses (chosen from 0.1, 0.3, 0.5, and 1.0 $M_\text{J}$), studied moons that were orbiting all of the planets out to distances of 1/2 of the Hill radius (compared to 1/7 for this study), and modeled the effects of oblateness.
%The systems considered in this paper do not cover the same breadth of initial conditions, but it is still useful to compare the two 

Comparing the moon survival rate in Figure~\ref{fig:moon_dist} to the corresponding results (\citet{Hong:2018}, Figure 5), we find a survival rate vs. distance relationship that is larger as a whole, and decays at a slower rate with distance.  Most of the differences between our results can result from the differences in our initial system properties.  Smaller mass perturbers will require more close encounters before a planet can be ejected---implying more scattering events and its accompanying potential for losing additional moons.  Moons on larger orbits are easier to ionize from the host planet.  And, lower-mass planets are likely to lose more moons when perturbed by larger counterparts.  %A portion of these differences are likely caused by the population of systems where all moons survive since they contribute to the survival rate at all distances equally.  If this additional population is removed, then our resulting distribution is consistent with the corresponding results in the other work.

\section{Conclusions}

%We have shown that moon systems that may exist during giant planet scattering are not completely disrupted, and have a chance of survival as their host planet is ejected from the system.  Our results predict a population of exomoons orbiting rogue gas giant planets.  These moons survive to a considerable distance from the host planet ($\gtrsim 200 R_{J}$).  The various encounters leading up to planetary escape leave the surviving moons in a wide range of orbital parameter space.  We see that precession, caused by planet-sun and planet-planet interactions, can excite the mutual inclinations of the moons in a system.

We have shown that giant planet scattering events will not completely disrupt moon systems even when the host planet is ejected from the system.  Our results predict a population of exomoons orbiting rogue gas giant planets.  These moons survive to a considerable distance from the host planet ($\gtrsim 200 R_{J}$).  The various encounters leading up to planetary escape leave the surviving moons in a wide range of orbital parameter space.  We see that precession, caused by planet-sun and planet-planet interactions, can excite the mutual inclinations of the moons in a system.  Nevertheless, the moon systems can often retain key dynamical properties.

%Some moons are removed from the planet will remain bound to the star system in a wide variety of heliocentric orbits.  Many of these moons are in smaller, eccentric orbits, with a significant fraction of moons occupying high inclinations.  Over long timescales such moons would likely interact with the remaining planets in the system and either be ejected, or collide with the central star or planets.  Some fraction, after additional scatterings that bring their pericenter distances closer to the star, may have their orbits decay through tidal dissipation---placing them on very short orbits and possibly contributing to the population of ultra short period planets \citep{Sanchis-Ojeda:2014,Steffen:2016}.

Some moons that are stripped from the planet will remain bound to the star in heliocentric orbits.  Many of these moons are in smaller, eccentric orbits, with a significant fraction of moons occupying high inclinations.  Over long timescales such moons would likely interact with the remaining planets in the system and either be ejected, or collide with the central star or planets.  Some fraction, after additional scatterings that bring their pericenter distances closer to the star, may have their orbits decay through tidal dissipation---placing them on very short orbits and possibly contributing to the population of ultra short period planets \citep{Sanchis-Ojeda:2014,Steffen:2016}.

%Although Jupiter-mass planets were used in these simulations, it is possible for similar scenarios to occur between planets of lower masses, such as Saturn or Neptune-mass planets.  Although a detailed study on the effect of planetary masses is beyond the scope of this project, we expect lower-mass planets to have a longer time to a planetary ejection, and possibly a smaller range in values for unbound moons due to the multiple encounters.  However, certain features, such as the inclination-precession behavior of bound moons in circular orbits, are expected to remain as the cause of these features is independent of the host planet mass.

Although we used Jupiter-mass planets in these simulations, similar scenarios can occur with lower mass planets, such as Saturn or Neptune-mass planets.  Although a detailed study on the effect of planetary masses is beyond the scope of this project, we expect lower-mass planets to require more encounters prior to a planetary ejection possibly producing a smaller range in values for remaining moons due to the multiple encounters.  Nevertheless, certain features, such as the inclination-precession behavior of bound moons in circular orbits, are expected to remain as the cause of these features is independent of the host planet mass.

%FIXME: JUST A NOTE.  YOU ORIGINALLY HAD THIS PARAGRAPH AS THE LAST ONE IN THE PAPER.  IN OTHER WORDS, WE SAID "WE JUST FOUND A NEW RESERVOIR FOR LIFE IN THE GALAXY THAT COULD CONSTITUTE BILLIONS OF POTENTIAL SYSTEMS" AND THEN WENT ON TO "NEPTUNE-LIKE PLANETS MAY PRODUCE DIFFERENT RESULTS FOR MOONS BECAUSE THEY ARE LESS MASSIVE".  I ALMOST THREW UP IN MY MOUTH.

%We find that resonant interactions are able to persist through planetary escape.  In our simulations, moon systems with a sustained resonance did not undergo large excitations to their eccentricities, and were able to remain coplanar despite the influences from the star and other planets.  These results, along with the survival rate shown in Figure~\ref{fig:moon_dist}, imply a significant fraction of escaping planets can retain a moon system with a previously established resonance.

We find that resonant interactions can persist through planetary escape.  In our simulations, moon systems that maintained their resonance behavior did not undergo large excitations to their eccentricities, and were able to remain coplanar despite the influences from the star and other planets.  These results, along with the survival rate shown in Figure~\ref{fig:moon_dist}, imply a significant fraction of escaping planets can retain a resonant moon system.

%How disturbed are the new moon systems?
%Ejection time relevant? Possibly not realistic planet formation.
An important consequence of our results is that moons like the Galilean satellites, which are geologically active and can sustain subterranean liquid water, could exist around rogue planets.  Moreover, since the mechanism that sustains the liquid water (tidal heating) can persist for billions of years, it raises the significant possibility for life to develop or persist around these planets in the absence of a host star.  Thus, lunar systems around rogue planets may constitute a new reservoir for life.  If a significant fraction of hot Jupiter systems are formed by planet-planet scattering (hot Jupiters exist around roughly 1\% of stars \citep{Wright:2012}), then the number of potential life-harboring rogue planets could be hundreds of millions or billions across the galaxy.

\section*{Acknowledgements}

Simulations in this paper made use of the REBOUND code, which can be downloaded freely at http:/github.com/hannorein/rebound. The author would like to thank Hanno Rein for his assistnace with some of the finer points of the REBOUND integrator.  We acknowledge support from NASA grants NNX16AK32G and NNX16AK08G.  We also thank Benjamin Bromley for useful discussions.

\bibliographystyle{mnras}
\bibliography{ref} % if your bibtex file is called example.bib

%\begin{thebibliography}{99}
%\bibitem[\protect\citeauthoryear{Others}{2013}]%{Others2013}
%\end{thebibliography}

%%%%%%%%%%%%%%%%%%%%%%%%%%%%%%%%%%%%%%%%%%%%%%%%%%

%%%%%%%%%%%%%%%%% APPENDICES %%%%%%%%%%%%%%%%%%%%%

%\appendix



%%%%%%%%%%%%%%%%%%%%%%%%%%%%%%%%%%%%%%%%%%%%%%%%%%

% Don't change these lines
\bsp	% typesetting comment
\label{lastpage}
\end{document}